\documentclass[twoside]{ilcws08}
\usepackage[latin1]{inputenc}
\usepackage[dvips]{graphicx,epsfig,color}
\usepackage{wrapfig,rotating}
\usepackage{amssymb,amsmath,array}
\usepackage{booktabs}
\begin{document}
\title{Feasibility study of Higgs pair creation in $\gamma\gamma$ collider}
\author{Keisuke Fujii$^1$, Katsumasa Ikematsu$^2$, Shinya Kanemura$^3$, \\
Yoshimasa Kurihara$^4$, Nozomi Maeda$^5$, Tohru Takahashi$^6$,
\vspace{.3cm}\\
1, 2, 4- KEK, Tsukuba, Japan \\
3- Department of Physics, University of Toyama, Japan \\
5, 6- Advanced Sciences of Matter, Hiroshima University, Higashi-Hiroshima, Japan}
\maketitle

\begin{abstract}
We studied a feasibility of measuring Higgs boson pair production in a Photon Linear Collider.  The optimum energy of $\gamma \gamma$ collision was estimated with a realistic luminosity distribution.  We also discussed simulation study for detecting the signal against W boson pair backgrounds.
\end{abstract}

\begin{wrapfigure}{r}{0.5\columnwidth}
\vspace{-0.8cm}
\centerline{\includegraphics[width = 0.45\columnwidth]{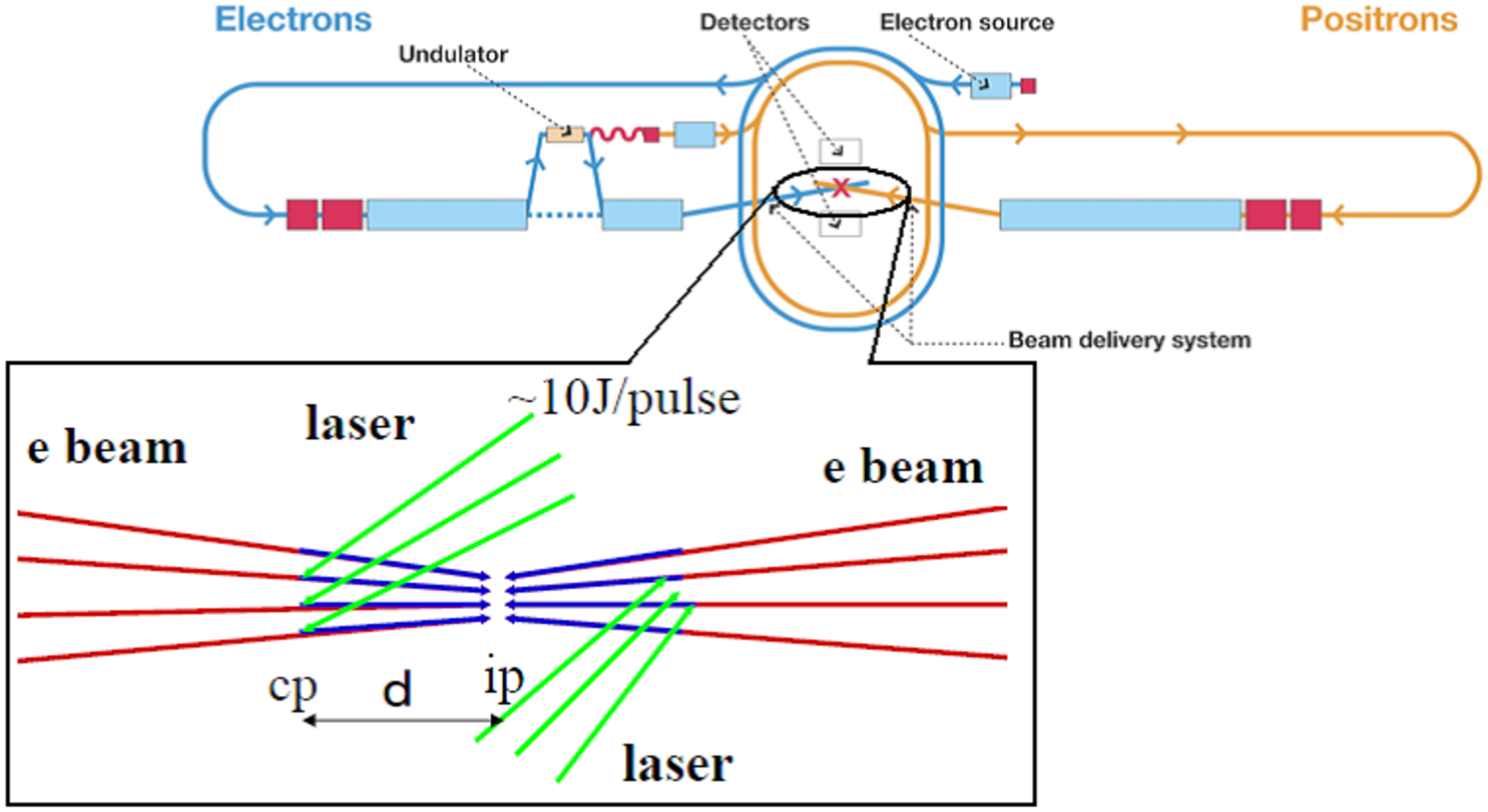}}
\vspace{-0.4cm}
\caption{An outline of PLC. Positron beam of ILC is replaced with electron beam. High
energy photon is generated by collision between laser and electron beam. \vspace{0.7cm}}
\label{PLC}

\centerline{\includegraphics[width = 0.45\columnwidth]{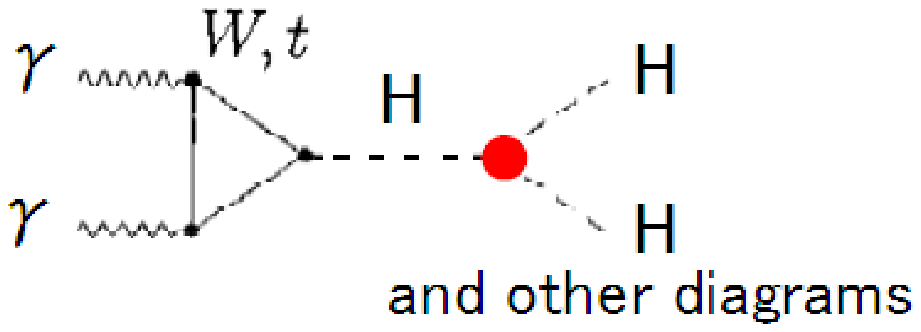}}
\vspace{-0.4cm}
\caption{An example of $\gamma\gamma \to HH$ diagram.  Higgs self-coupling occurs at red point.\vspace{0.7cm}}
\label{AAHH}

\centerline{\includegraphics[width=0.30\columnwidth]{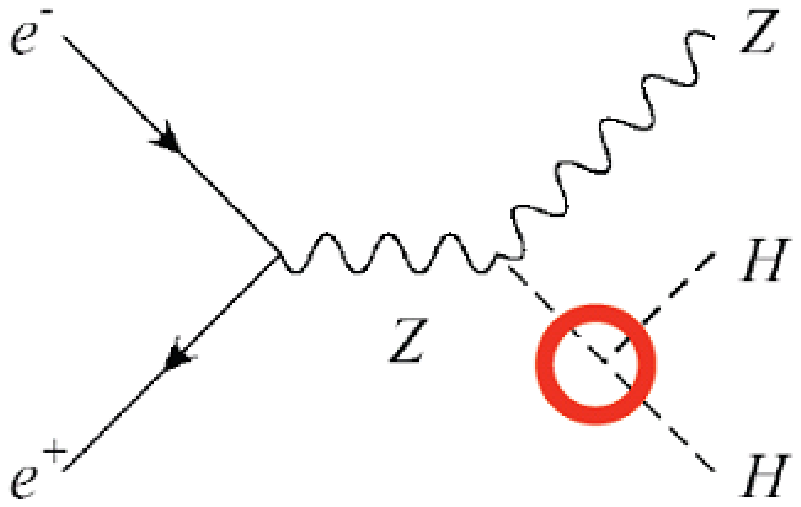}}
\vspace{-0.4cm}
\caption{$e^+e^- \rightarrow ZHH$ diagram.  Higgs self-coupling occurs at red circle.}
\vspace{-0.5cm}
\label{eeZHH}
\end{wrapfigure}

\section{Introduction}
As a possible option of the International Linear Collider, feasibility of physics orotundities of high energy photon-photon interaction has been considered.  In the high energy photon linear colliders(PLCs), high energy photon beams are generated by inverse Compton scattering between the electron and the laser beams as illustrated in figure \ref{PLC}.  Feasibility of the PLC for both physics and technical aspect, has been studied and summarized in \cite{phys.summary}.  In these study, one assumed integrated luminosity of 3~4 years PLC operation which, for example, may happens after initial operation of $e^+ e^-$ mode of the ILC at $\sqrt{s} = 500$GeV.

In this study, we investigated a feasibility of self-coupling of the Higgs boson as an example of a precise measurement with the PLC by assuming an ultimate integrated luminosity, i.e., 10years operation with a high luminosity parameters.

The Higgs boson self-coupling constant is reprensented by $\lambda = \lambda^{SM} (1+\delta\kappa)$ which contributes Higgs boson pair production via a diagram shown in figure \ref{AAHH}.  Here, $\lambda^{SM}$ is Higgs boson self-coupling constant which is included in the Standard Model.  $\delta\kappa$ represent the deviation from the Standard Model.

The self-coupling of the Higgs boson can also be studied in $e^+e^-$ collision via the diagram shown in figure \ref{eeZHH}.  Comparing with the $e^+e^- \to ZHH$ channel, where Higgs boson pairs are associated with the Z boson production, the Higgs bosons are produced by s channel via loop diagrams in $\gamma \gamma$ collision.  Therefore, contribution of the $\delta\kappa$ to the production cross section is difference for the $e^+e^-$ and for $\gamma \gamma$ and studies in these two modes will be complementary each other.  Detail of the theoritical background in this analysis can be found in \cite{DetailBG}.

\section{Sensitivity Sutdy}
For optimization photon-photon collision energy, we defined the sensitivity for the $\delta\kappa$ as;

\begin{equation*}
sensitivity = \frac{|N(\delta\kappa) - N_{SM}|}{\sqrt{N_{obs}}}
= \frac{L|\eta \sigma (\delta\kappa) - \eta \sigma_{SM}|}{\sqrt{L( \eta \sigma (\delta\kappa) + \eta_B \sigma_B)}}
\end{equation*}

\begin{wrapfigure}{r}{0.5\columnwidth}
\vspace{-0.3cm}
\centerline{\includegraphics[width=0.45\columnwidth]{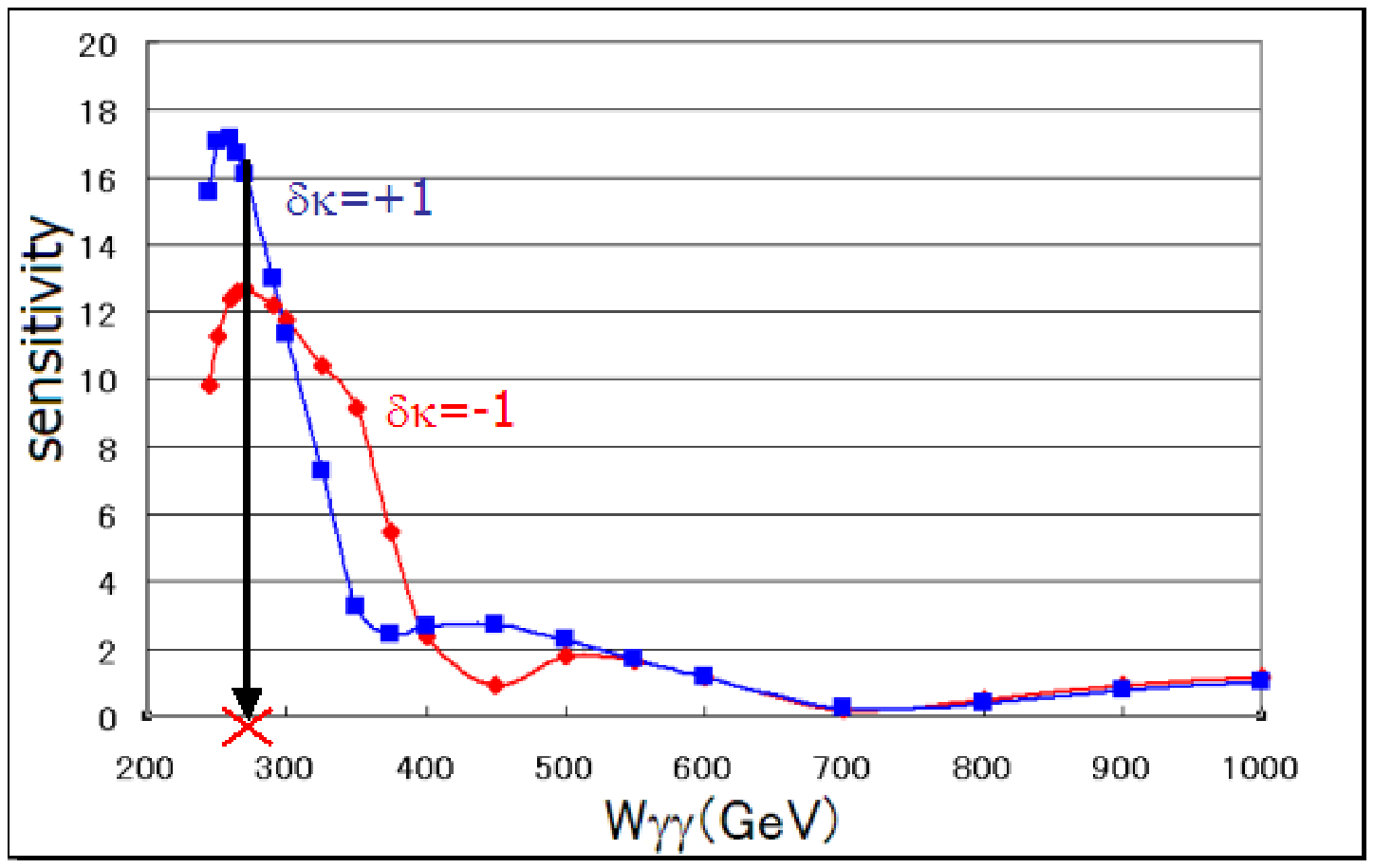}}
\vspace{-0.4cm}
\caption{A graph showing sensitivity v.s. $W_{\gamma \gamma}$.  $W_{\gamma \gamma}$ means photon-photon collision energy.  Sensitivity has peak at near $W_ {\gamma \gamma} \simeq 270$GeV, not depend on $\delta \kappa$.}
\label{sensitivity}

\vspace{0.5cm}
\centerline{\includegraphics[width=0.42\columnwidth ]{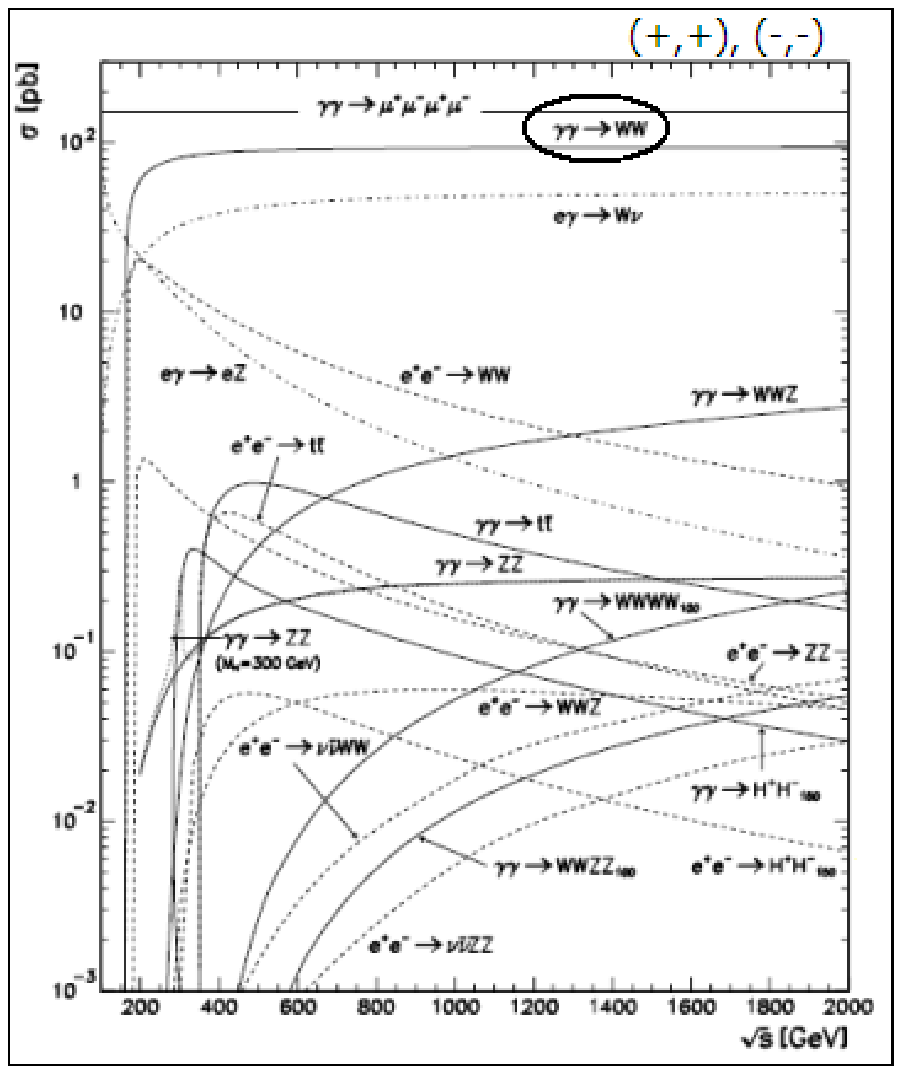}}
\vspace{-0.4cm}
\caption{cross section v.s.\,collision energy.  $\gamma \gamma \to WW$ is main background against $\gamma\gamma \to HH$.}
\vspace{-1.7cm}
\label{crosssection}
\end{wrapfigure}

\hspace{-16pt}where, $N(\delta \kappa)$ is a expected number of events as a function of $\delta\kappa$ and $N_{SM}$ is the number of events expected from the Standard Model.  $L, \eta, \sigma(\delta\kappa), \sigma_{SM}, \eta_B$ and $\sigma_B$ are integrated luminosity, detection efficiency of signal, cross section with $\delta\kappa$, cross section with the Standard model, detection efficiency for background events and the cross section for background  processes, respectively.  For $\eta = 1$, $\eta_B = 0$, sensitivity is written;
\begin{equation*}
sensitivity = \sqrt{L} \frac{|\sigma(\delta\kappa) - \sigma_{SM}|}{\sqrt{\sigma (\delta \kappa)}}
\end{equation*}

\hspace{-14pt}The Higgs boson mass of 120GeV and the integrated luminosity of 1000fb$^{-1}$ was assumed in the study.  The cross section is calculated by the formula which is described in \cite{Xsecfomula} with a theoretically calculated PLC luminosity spectrum.  The sensitivity as a function of the center of mass energy of the $\gamma \gamma$ collision for $\delta \kappa = $ 1 and -1 is plotted in figure \ref{sensitivity}. 

From the figure, the optimum energy for the $\gamma \gamma$ collision for Higgs boson mass of 120GeV was found to be around 270GeV.

\section{Background}
Figure \ref{crosssection} shows cross section as a function of collision energy for photon-photon interactions.  Figure \ref{crosssection} indicates that $\gamma\gamma \to WW$ is main background with the production cross section of about 90pb.  On the other hand, 
\begin{wraptable}{r}{0.5\columnwidth}
\vspace{-0.4cm}
\caption{Input parameters to CAIN.  This parameters set make luminosity peak at optimum energy.}
\vspace{-0.2cm}
\label{parameters}
 \begin{center}
 \begin{tabular}{lr} \toprule
 $E_e[GeV]$ & 190\\ \hline
 N/$10^{10}$ & 2 \\ \hline
 $\sigma_z[mm]$ & 0.35 \\ \hline
 $\gamma \varepsilon_{x/y}/10^{-6}[mrad]$ & 2.5/0.03 \\ \hline
 $\beta_{x/y}[mm]@IP$ & 1.5/0.3 \\ \hline
 $\sigma_{x/y}[nm]$ & 96/4.7 \\ \hline
 $\lambda_L[nm]$ & 1054 \\ \hline
 Pulse energy$[J]$ & 10 \\ \hline
 $x = 4\omega E_e /m_e^2$ & 3.76 \\ \bottomrule
 \end{tabular}
 \end{center}
 \vspace{-0.2cm}
\end{wraptable}

\hspace{-14pt}signal cross section is 0.044fb at optimized energy.  Therefore, observation of signal requires background suppression of $10^{-7}$.  The other reaction that has large cross section such as $\gamma\gamma \to WWZ$ and $\gamma \gamma \to t\bar{t}$.  However the optimum energy for $\gamma \gamma \to HH$ is below these threshold for these channel.
\begin{wrapfigure}{r}{0.5\columnwidth}
\vspace{-2.3cm}
\centerline{\includegraphics[width=0.45\columnwidth]{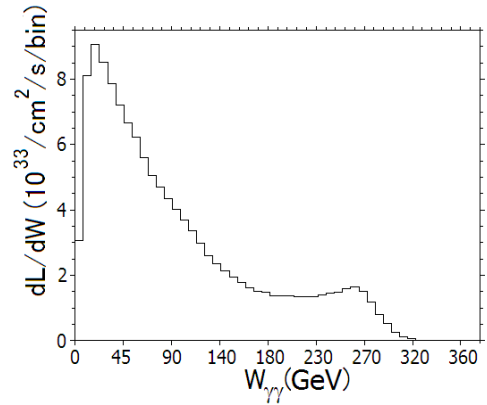}}
\vspace{-0.4cm}
\caption{Luminosity spectrum generated by CAIN using table \ref{parameters} parameters set.}
\label{Luminosity}
\vspace{-0.5cm}
\end{wrapfigure}

\section{Simulation Framework}
JLC Study Framework(JSF) is used as simulation framework in this study \cite{JSF}.

The helicity amplitude for the signal is calculated by theoritical calculation program \cite{Kanemura}.  The helicity amplitude for background processes were calculated by a helicity amplitude calculation package; HELAS \cite{HELAS}.

The luminosity distribution used in the analysis were generated using the CAIN\cite{CAIN} program with the input parameters shown in table \ref{parameters} \cite{CAINinput}.  The luminosity spectrum simulated with the CAIN is shown in figure \ref{Luminosity}.

From these helicity amplitude and luminosity spectrum, BASES/SPRING integrated and generated events.  Pythia made parton shower and hadronized.  Quick detector simulator read particle list that from pythia.  Finaly, data from Quick Detector Simulator is analyzed.

\begin{wraptable}{r}{0.5\columnwidth}
\vspace{-0.8cm}
\caption{Branching ratio of Higgs particle.}
\vspace{-0.2cm}
\label{BRH}
 \begin{center}
 \begin{tabular}{cr} \toprule
 particles & Branching ratio \\ \midrule
 $b\bar{b}$ & 0.6774  \\ \midrule
 $\mu \mu $ & 0.00024  \\ \midrule
 $c\bar{c}$ & 0.02982 \\ \midrule
 $\tau \tau$ & 0.06916 \\ \midrule
 $s\bar{s}$ & 0.00051 \\ \midrule
 $gg$ & 0.0713 \\ \midrule
 $\gamma \gamma$ & 0.002231 \\ \midrule
 $\gamma Z$ & 0.001084 \\ \midrule
 $WW$ & 0.1331 \\ \midrule
 $ZZ$ & 0.0152 \\ \bottomrule
 \end{tabular}
 \end{center}
\vspace{-1.8cm}
\end{wraptable}

With this spectrum, we expect signal of 16event/year, while $10^7$event/year for background.

\section{Analysis}
The decay branching ratio of the Higgs boson of 120GeV is shown in table \ref{BRH}.  Since main decay mode of the Higgs boson of 120GeV is b-quark pairs with the branching ratio of about 0.67, we tried the case that both Higgs boson decayed into b-quark pairs.

\begin{wrapfigure}{r}{0.5\columnwidth}
\vspace{-1.5cm}
\centerline{\includegraphics[width=0.42\columnwidth]{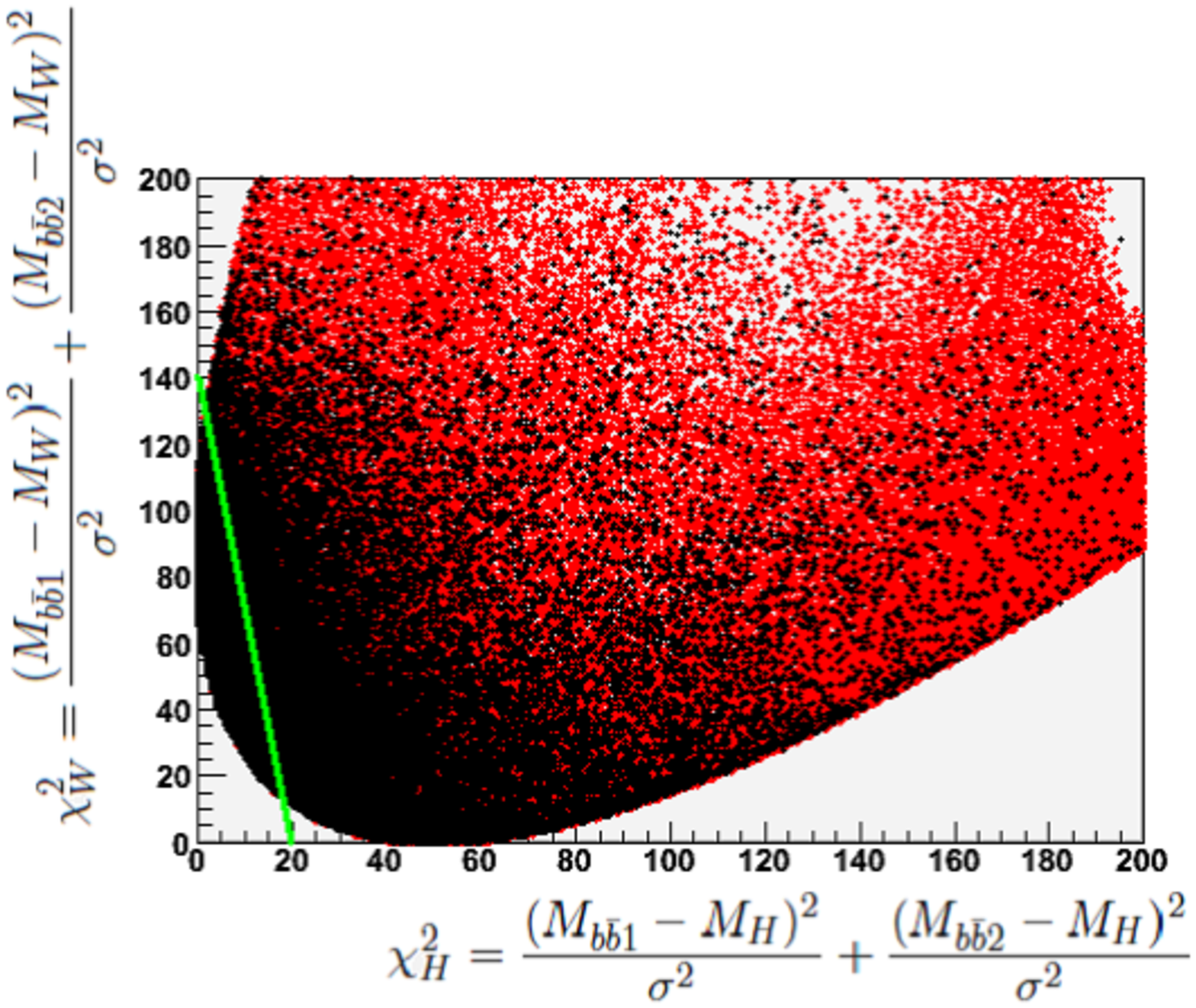}}
\vspace{-0.4cm}
\caption{Reconstructed particles $\chi^2$ distribution.  Black indicates signal events, red indicates background events.  Green line is represented by $-140/20 \times \chi_H^2 + 140 = \chi^2_W$.  Here, to make signal clear, signal cross section is about $5 \times 10^4$ times as large as usual.}
\label{chi2}

\vspace{0.5cm}
\centerline{\includegraphics[width=0.42\columnwidth]{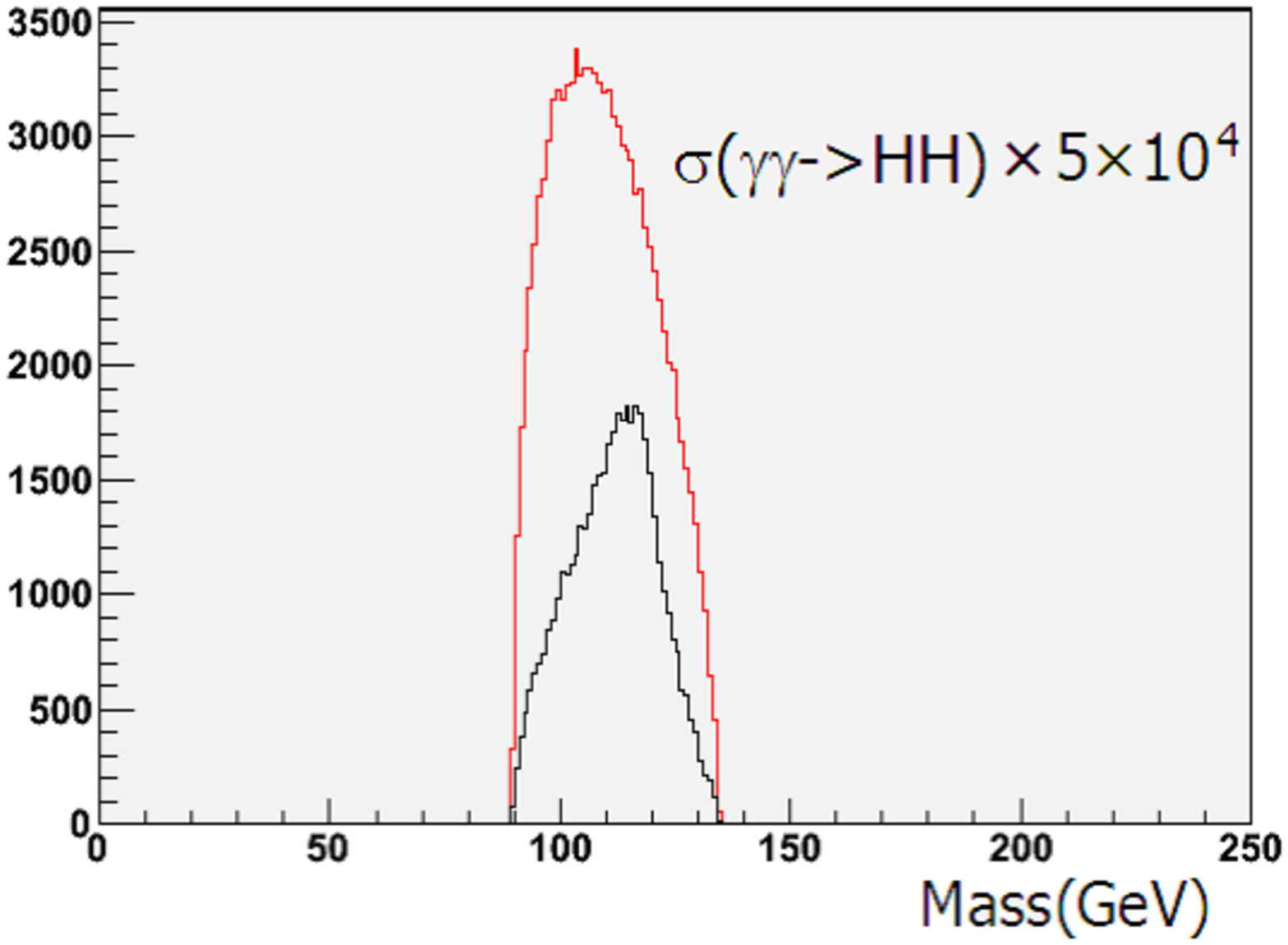}}
\vspace{-0.4cm}
\caption{Reconstructed particle mass spectrum that cutted.  Background is suppressed, but not enough.}
\label{masscuted}
\vspace{0.7cm}

\centerline{\includegraphics[width=0.49\columnwidth]{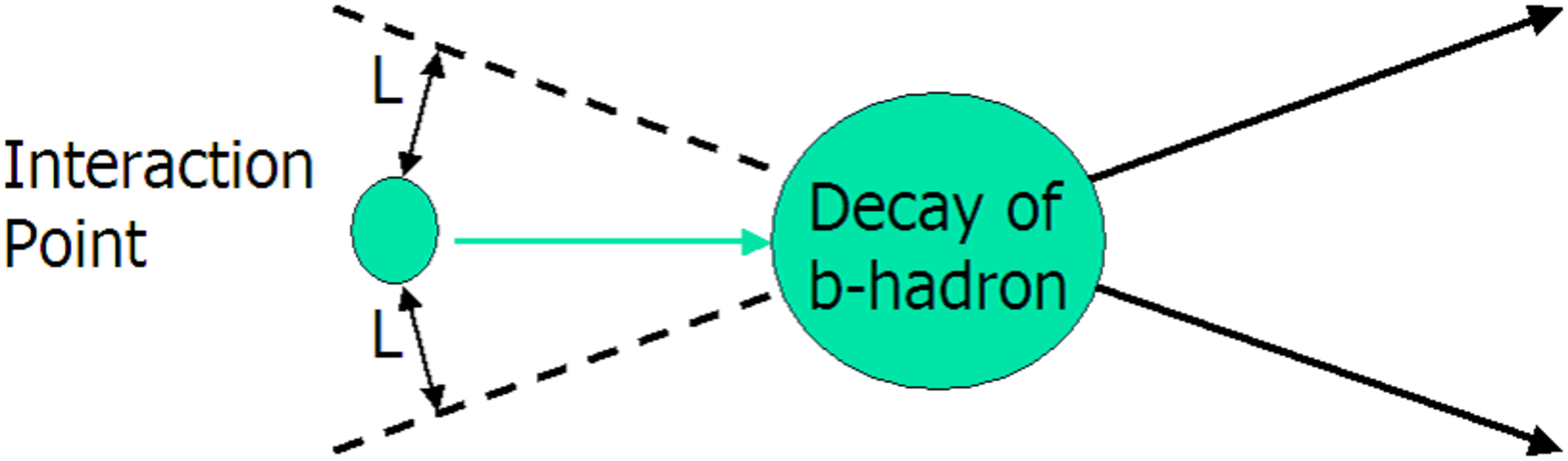}}
\vspace{-0.4cm}
\caption{An outline of nsig method.  B-hadron is generated at interaction point and decay at "Decay of b-hadron".  Arrows mean particle tracks.  Dotted line means extrapolate particle tracks.}
\label{nsig}
\vspace{-3.0cm}
\end{wrapfigure}

For each event, we applied forced four jets analysis in which a clustering algorithm is applied to an event by changing the clustering parameter until the event is categorized as a four jets event.  After the forced four jets analysis, invariant masses for jet pairs were calculated.  For a four jets event, we must to choose a right jets pairs originating from parent Higgs (or W for the background) bosons out of three possible 
combinations.  For this purpose, we defined $\chi^2$s as;

\hspace{-1cm}
\begin{minipage}{0.5\columnwidth}
\begin{gather}
\chi_H^2 = \frac{(M_1 - M_H)^2}{\sigma_{2j}^2} + \frac{(M_2 - M_H)^2}{\sigma _{2jH}^2} \notag \\
\chi_W^2 = \frac{(M_1 - M_W)^2}{\sigma_{2j}^2} + \frac{(M_2 - M_W)^2}{\sigma _{2jW}^2} \notag
\end{gather}
\end{minipage}

\hspace{-15pt}where, $M_1$ and $M_2$ are reconstructed particle mass, $M_H$ and $M_W$ are Higgs boson and W boson mass respectively, with $\sigma_{2jH}$ and $\sigma_{2jW}$ being their resolutions.
The jet of the least $\chi^2$ was chosen to be the most probable combination for an event.  Figure \ref{chi2} shows correlation of $\chi_H^2$ and $\chi_W^2$ for the most probable combination.  To enhance Higgs boson from the W boson events, we choosed an event satisfies $-140/20 \times \chi_H^2 + 140 \geq \chi^2_W$.  The mass distributions for the Higgs and W boson events after $\chi^2$ cut are shown in figure \ref{masscuted}.

\section{b-tagging}
By the $\chi^2$ analysis described in previous section, the W boson background was suppressed by 0.0541 while keeping the 46$\verb|%|$ efficiency for the Higgs boson events.  In order further improve signal to background ratio, we applied b-quark tagging method for remaing events.

\begin{wrapfigure}{r}{0.5\columnwidth}
\vspace{-0cm}
\centerline{\includegraphics[width=0.42\columnwidth]{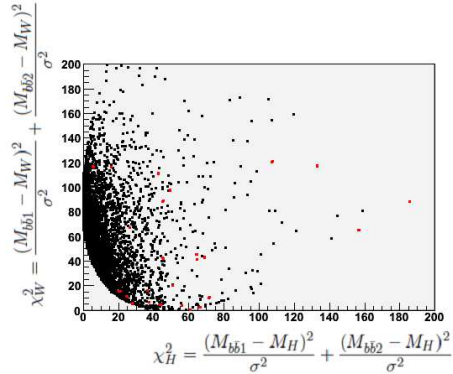}}
\vspace{-0.4cm}
\caption{A result of b-tag selection.  Number of b-tagged jets = 4 is required.  Black indicates signal event.  Red indicates background event.  Number of remained background event is 52.}
\label{chi2btagged}
\vspace{-1.1cm}
\end{wrapfigure}

Figure \ref{nsig} illustrates a b-quark tagging method we applied.  For each track in a reconstucted jet, $N_{sig} = L/\sigma_L$ was calculated, where L is the least approach to the interaction point of the track in the plane perpendicular to the beam and $\sigma_L$ being its resolution.  Then, $N_{off}(a)$, number of track which has $N_{sig} > a$, is calculated for each jet as a function of a.  In current analysis we requied all jets must satisfy $N_{off}(3.5) \geq 2$.  Figure \ref{chi2btagged} is the $\chi^2$ plot after b-tagging but before $\chi^2$ cuts.  We obtaned backgroud suppression of $1.35 \pm 0.18 \times 10^{-6}$ and efficiency of signal of $0.1454 \pm 0.0044$, where the errors 

\clearpage

\hspace{-15pt}are from statistic of the Monte Carlo simulation.  For remaining envents, $\chi^2$ cut were applyied.  As a results, no WW events survied out of $3.85 \times 10^7$ simulated events while keeping signal efficiency of $0.1096 \pm 0.0014$.

\section{Summary and prospect}
We studied feasibility of measurement of Higgs self-coupling constant at the PLC.  For Higgs mass of 120GeV, optimum photon-photon collision energy for observe $\gamma \gamma \to HH$ was found to be about 270GeV.  With a parameters of PLC(TESLA-optimistic), 16events/year is expected for Higgs boson events while main background of $\gamma \gamma \to WW$ is about $10^7$events/year.

We tried an event selection with kinematical parameters and b-quark tagging by the simulation and found that backgound suppression of $10^{-7}$ with keeping signal efficiency of about $10\verb|%|$ seemed to be possible.

\hspace{-15pt}For further analysis, we plan to improve signal efficicency by :

optimization of selection criteria for $HH \to b\bar{b} b\bar{b}$ mode.

study for $HH \to b\bar{b}WW^*$ decay.

For the backgound, it is necessary to estimate contribution from $\gamma \gamma \to ZZ$ events.

\section{Acknowledgement}
The authors would like to thank the ILC physics working group for valuable discussion and suggestion.


\begin{thebibliography}{9}
\bibitem{phys.summary}
E\verb|.|Boos, A\verb|.|De Roeck, I\verb|.|F\verb|.|Ginzburg,  K\verb|.|Hagiwara, R\verb|.|D\verb|.|Heuer, G\verb|.|Jikia, J\verb|.|Kwiecinski, D\verb|.|J\verb|.|Miller, T\verb|.|Takahashi, V\verb|.|I\verb|.|Telnov, T\verb|.|Rizzo, I\verb|.|Watanabe, P\verb|.|M\verb|.|Zerwas
\textit{"Gold-plated processes at photon colliders"} (2001) arXiv:hep-ph/0103090v1

\bibitem{DetailBG}
G\verb|.|Jikia
\textit{"Pair production of W bosons at the photon linear collider:a window to the electroweak symmetry breaking?"} (1997) arXiv:hep-ph/9708373v1

\bibitem{Xsecfomula}
E\verb|.|Asakawa, D\verb|.|Harada, S\verb|.|Kanemura, Y\verb|.|Okada, K\verb|.|Tsumura
\textit{"Higgs boson pair production at a photon-photon collision in the two Higgs doublet model"} (2008) arXiv:0809.0094v2[hep-ph]

\bibitem{JSF}
\textit{"The JSF home page"} http://www-jlc.kek.jp/subg/offl/jsf/

\bibitem{Kanemura}
\textit{"Shinya Kanemura's page"} http://jodo.sci.u-toyama.ac.jp/$\verb|~|$kanemu/

\bibitem{HELAS}
\textit{"HELAS Tutorial"} http://madgraph.kek.jp/$\verb|~|$kanzaki/Tutorial/tutorial.html

\bibitem{CAIN}
\textit{"Available computer programs on FFIR"} http://www-jlc.kek.jp/subg/ir/Program-e.html

\bibitem{CAINinput}
IB\verb|.|Badelek, et al.,
\textit{"The Photon Collider at TESLA"} (2004) International Journal of Modern Physics A19/, 5097-5186




\end{thebibliography}
\end{document}